\pdfoutput=1 

\documentclass[12pt]{article}

\usepackage{scicite}

\usepackage{times}
\usepackage{graphicx}
\usepackage{pdfpages}

\newcommand{\aver}[1]{\ensuremath{\langle {#1} \rangle}}
\newcommand{\ket}[1]{\left|  #1 \right\rangle}

\newenvironment{sciabstract}{%
\begin{quote} \bf}
{\end{quote}}

\newcounter{lastnote}
\newenvironment{scilastnote}{%
\setcounter{lastnote}{\value{enumiv}}%
\addtocounter{lastnote}{+1}%
\begin{list}%
{\arabic{lastnote}.}
{\setlength{\leftmargin}{.22in}}
{\setlength{\labelsep}{.5em}}}
{\end{list}}

\title{All-Optical Switch and Transistor Gated by One Stored Photon}

\author
{Wenlan Chen$^{1}$, Kristin M. Beck$^{1}$, Robert B\"{u}cker$^{1,2}$, \\ Michael Gullans$^{3}$, Mikhail D. Lukin$^3$, Haruka Tanji-Suzuki$^{1,3,4}$, \\ Vladan Vuleti\'{c}$^{1*}$\\
\normalsize{$^{1}$ Department of Physics and Research Laboratory of Electronics, }\\
\normalsize{Massachusetts Institute of Technology, Cambridge, Massachusetts 02139, USA}\\
\normalsize{$^{2}$ Vienna Center for Quantum Science and Technology, }\\
\normalsize{Atominstitut, TU Wien, Stadionallee 2, 1020 Vienna, Austria}\\
\normalsize{$^{3}$ Department of Physics, Harvard University, Cambridge, Massachusetts 02138, USA}\\
\normalsize{$^{4}$ Photon Science Center, School of Engineering,}\\
\normalsize{The University of Tokyo, 2-11-16 Yayoi, Bunkyo-ku, Tokyo 113-8656, Japan}\\
\normalsize{$^\ast$To whom correspondence should be addressed; E-mail: vuletic@mit.edu.}
}

\date{}

\begin{document}

% Double-space the manuscript.

\baselineskip24pt

% Make the title.

\maketitle

\baselineskip12pt
\textit{\centering This is the author's version of the work.\\It is posted here by permission of the AAAS for personal use, not for redistribution. \\The definitive version was published in Science Vol. 341 (16 August 2013), \\doi:10.1126/science.1238169 }

\baselineskip 24pt

% Place your abstract within the special {sciabstract} environment.

\begin{sciabstract}
The realization of an all-optical transistor where one `gate' photon controls a `source' light beam, is a long-standing goal in optics. By stopping a light pulse in an atomic ensemble contained inside an optical resonator, we realize a device in which one stored gate photon controls the resonator transmission of subsequently applied source photons. A weak gate pulse induces bimodal transmission distribution, corresponding to zero and one gate photons. One stored gate photon produces fivefold source attenuation, and can be retrieved from the atomic ensemble after switching more than one source photon. Without retrieval, one stored gate photon can switch several hundred source photons. With improved storage and retrieval efficiency, our work may enable various new applications, including photonic quantum gates, and deterministic multiphoton entanglement.
\end{sciabstract}

Photons are excellent carriers of quantum information, but it is difficult to induce the strong interactions between individual photons that are required for, e.g., all-optical quantum information processing. Nevertheless, advances toward such interactions have been made in cavity quantum electrodynamics (QED) systems with atoms \cite{Birnbaum05,Brennecke07,Colombe07,Kubanek08,Tanji-Suzuki11,Brooks12} or artificial atoms \cite{Michler00,Press07,Fushman08,Volz12,Bose12}, and in a cavity-free system using atomic Rydberg states \cite{Dudin12,Peyronel12} or dye molecules \cite{Hwang09}. All-optical switching of one beam by another\cite{Bajcsy09} and cross-phase modulation\cite{Lo11} have been demonstrated at the level of a few hundred photons by means of electromagnetically induced transparency (EIT)\cite{Schmidt96,Imamoglu97,Harris98,Fleischhauer00,Fleischhauer05}. At the few-photon level, nonclassical light has been generated \cite{Michler00,Birnbaum05,Press07,Kubanek08,Dayan08,Fushman08,Bose12,Dudin12,Peyronel12,Brooks12}, and optical nonlinearities of $16^\circ$ in phase shift\cite{Turchette95} and up to $\sim$20\% in two-photon attenuation\cite{Fushman08,Tanji-Suzuki11,Volz12} have been observed in cavity QED systems. While switching of the cavity transmission by a single atom has also been achieved \cite{Thompson92}, the realization of an optical transistor exhibiting gain with gate signals at the few- or one-photon level \cite{Chang07} remains a challenge.

We demonstrate a cavity QED version \cite{Imamoglu97} of an optical switch  \cite{Chang07} based on EIT in a four-level system \cite{Schmidt96,Imamoglu97,Harris98} where the collective atomic excitation associated with the storage of one gate photon \cite{Fleischhauer00,Liu01,Phillips01} blocks the resonator transmission. Our system \cite{Tanji-Suzuki11} consists of an ensemble of laser-cooled cesium atoms optically trapped inside a high-finesse optical cavity (Fig.~1A) operating in the strong-coupling regime \cite{Birnbaum05,Brennecke07,Colombe07,Kubanek08,Tanji-Suzuki11,Brooks12} of cavity QED. Each atom has a four-state $N$-type level structure $\ket{g} \leftrightarrow \ket{d} \leftrightarrow \ket{s} \leftrightarrow \ket{e}$ with two stable ground states $\ket{g}$, $\ket{s}$, and two electronic excited states $\ket{d}$, $\ket{e}$ (Fig.~1B). For atoms prepared in state $\ket{g}$, this atomic structure mediates an effective interaction between free-space photons (photons resonant with the $\ket{g} \rightarrow \ket{d}$ transition serving as gate photons) and cavity photons (photons resonant with the $\ket{s} \rightarrow \ket{e}$ transition serving as the source)\cite{Schmidt96,Imamoglu97,Harris98}. These two transitions are connected via a control laser that addresses the $\ket{d} \rightarrow \ket{s}$ transition and induces transparency (EIT) for the gate photons. By ramping the control laser power down to zero, we store a weak gate pulse inside the atomic ensemble (Fig.~1B), and retrieve it at a later time by adiabatically re-applying the control beam (Fig.~1D) \cite{Fleischhauer00,Liu01,Phillips01}. In between storage and retrieval, we apply a source beam (Fig.~1C). The atomic population in state $\ket{s}$ associated with the stored gate pulse can block the transmission of the source pulse through the cavity \cite{Thompson92}. Due to the finite optical depth of the ensemble ($OD \leq 0.9$) and sub-optimal control waveform \cite{Gorshkov2007}, one out of 5 to 10 incident gate photons is stored.

\begin{figure}[ht!]
  \centering
    \includegraphics{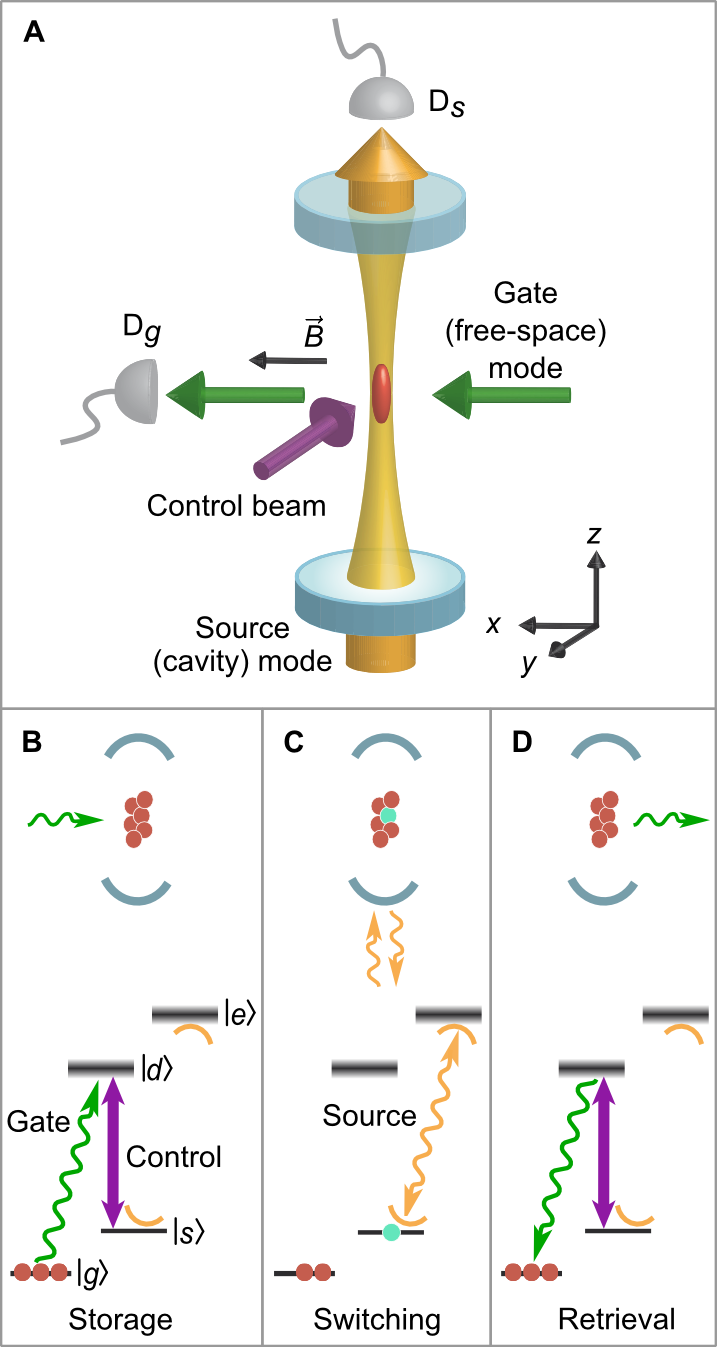}
  \caption{ {\bf Fig. 1. All-optical switch and transistor.} Setup (A) and atomic level scheme with experimental sequence (B-D). An ensemble of laser cooled atoms is trapped inside an optical resonator operating in the single-atom strong-coupling regime on the $\ket{s} \rightarrow \ket{e}$ transition. (B) We first store a gate photon in the medium, which corresponds to a collective atomic excitation to state $\ket{s}$. (C) This collective excitation blocks the transmission of source photons through the cavity and (D) can be retrieved. Retrieved gate and transmitted source photons are measured with photon counters $D_g$ and $D_s$, respectively. The atomic states of $^{133}$Cs used in this experiment are $\ket{g}=\ket{6S_{1/2},F=3,m_F=3}, \ket{d}=\ket{6P_{3/2},4,4}, \ket{s}=\ket{6S_{1/2},4,4}, \ket{e}=\ket{6P_{3/2},5,5}$, where $F$ and $m_F$ denote the hyperfine and magnetic sublevels. }
\end{figure}

We first characterize the cavity transmission without gate photon retrieval. To this end, we measure the average cavity transmission spectrum for different mean stored gate photon numbers $\aver{n_g}$ (Fig.~2). Since the gate pulses are weak classical pulses (coherent states), they are associated with Poissonian distributions in photon number $n_g$, and there is a finite probability $p(0)=e^{-\aver{n_g}}$ that the stored gate pulse does not contain any photons. Therefore, even if one photon were to perfectly switch off the source beam, there is a maximum average switching contrast $1-e^{-\aver{n_g}}$ for measurements with coherent states of gate photons (solid line in the inset to Fig.~2). The measured data points lie close to the maximum possible switching contrast, and within the theoretically expected range (gray area).

\begin{figure}[ht!]
   \centering
      \includegraphics{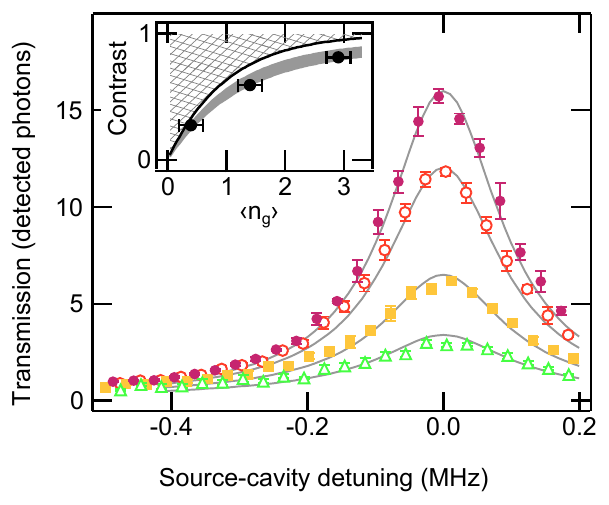}
   \caption{{\bf Fig. 2. Cavity transmission in the presence of stored gate photons.} Average transmission spectra of a source beam applied for $24 \mu$s for mean stored gate photon numbers $\aver{n_g}=0,0.4,1.4,2.9$ (top to bottom).  The solid lines are theoretical curves \cite{SOM}. Error bars are standard errors of the mean. The inset shows the relative transmission on cavity resonance (switching contrast) versus $\aver{n_g}$. The gray area indicates the theoretical prediction. The solid black line corresponds to the maximum average switching contrast that can be observed with coherent states of gate photons.}
\end{figure}

The photon number quantization of the gate pulse and the cavity blocking by just one gate photon are evident when we plot histograms of transmission spectra (Fig.~3) instead of the average transmission. The histogram shows two clearly separated components (Fig.~3B), where the high-transmission component corresponds to $n_g=0$, while the low-transmission component corresponds to $n_g \geq 1$ (mostly $n_g = 1$ gate photons. The high-to-low peak transmission ratio gives an extinction factor for one stored gate photon of $T^{-1}=11 \pm 1$.

\begin{figure}[ht!]
   \centering
      \includegraphics{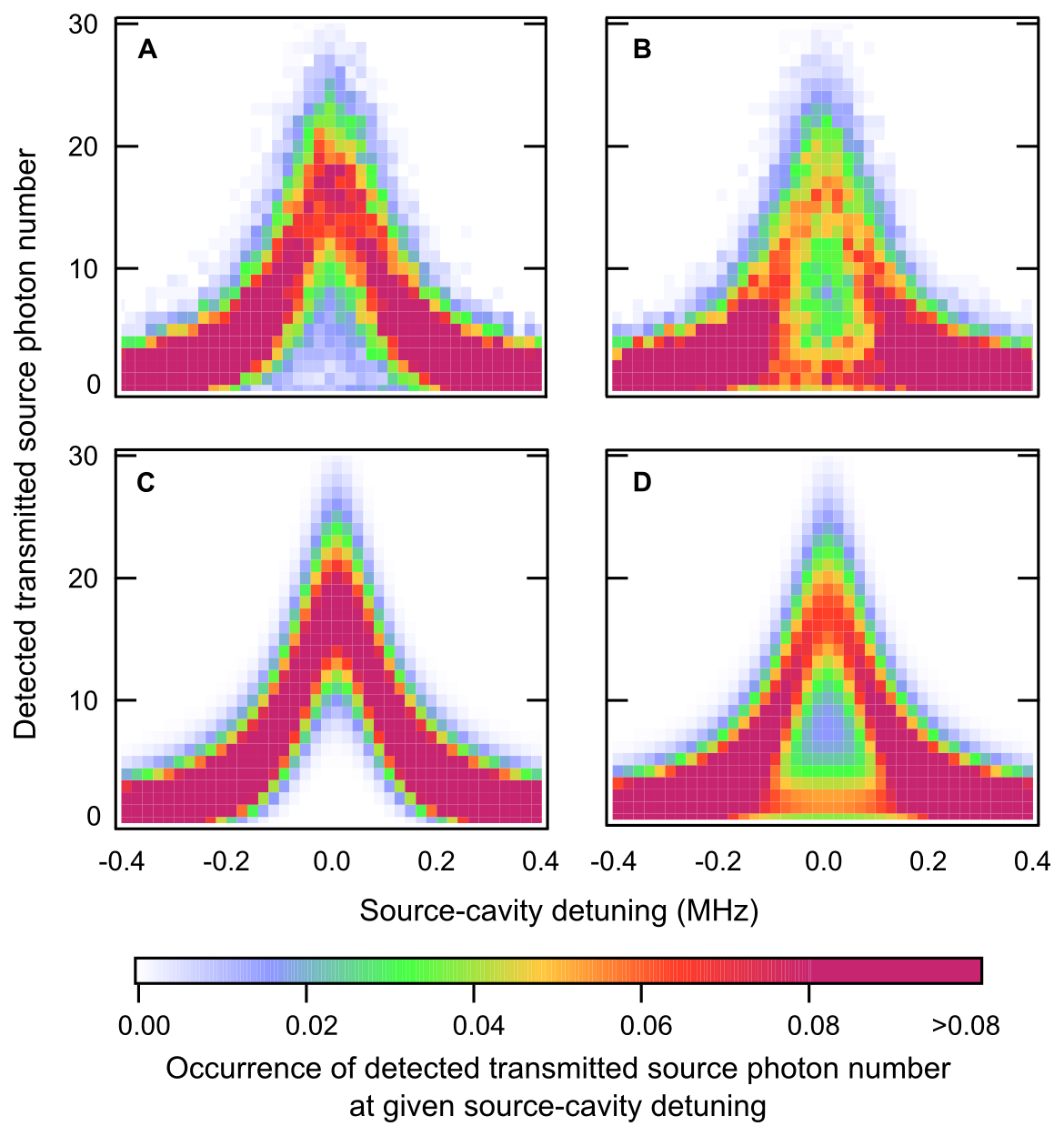}
   \caption{{\bf  Fig. 3. Histogram of cavity transmission spectra.} Cavity transmission without (A) and with $\aver{n_g}=0.5$ gate photons (B). The horizontal axis indicates the detuning of the source beam from the cavity resonance, the vertical axis the number of detected transmitted source photons in a $24~\mu$s detection window. The color indicates the occurrence rate of a particular detected transmitted source photon number for a given source-cavity detuning.  The histogram displays a clear separation between the zero-gate-photon component $n_g=0$ with high cavity transmission (17 detected source photons), and the component $n_g \geq 1$ ($n_g = 1$ with probability 0.8, $n_g>1$ with 0.2) leading to cavity blocking (1.5 detected photons). (C), (D) show the corresponding theoretically expected histograms. The extinction factor 17/1.5 for one gate photon is $T^{-1}=11 \pm 1$.}
\end{figure}

In order to characterize the optical gain of the system, we measure the distribution of the transmitted source photon number, $M_s= \frac{\mathcal{T}}{\mathcal{T}+\mathcal{L}} \int dt~ m_c(t) \kappa$, on cavity resonance. Here $m_c(t)$ is the intracavity photon number at time $t$, $\kappa$ is the cavity linewidth, and $\frac{\mathcal{T}}{\mathcal{T} + \mathcal{L}} = 0.66$ (with cavity mirror transmission $\mathcal{T}$ and mirror loss $\mathcal{L}$) accounts for the outcoupling efficiency of an intracavity photon. $M_s$ can be determined from the detected photon number and the independently measured detection-path efficiency \cite{SOM}. As Fig.~4A shows, the distribution is double peaked, with the high-transmission peak with average source photon number $\aver{M_s}\big{|}_{n_g=0}$ corresponding no gate photon, while the gray area of low transmission $\aver{M_s}\big{|}_{n_g\geq1}$ corresponds to the blocking by one or more gate photons. The optical gain per stored gate photon can then be defined as the gate-photon-induced change in source transmission, $G=\aver{M_s}\big{|}_{n_g=0}-\aver{M_s}\big{|}_{n_g\geq1}$, which is directly determined from the measured histogram. Fig.~4B shows the measured gain as a function of the applied source photon number, where the gain saturation occurring around 1000 source photons is likely due to optical pumping of the atom into magnetic sublevels with weaker cavity coupling. Remarkably, one stored gate photon can block more than $\sim600$ source photons, of which $\sim400$ are available outside the cavity.

\begin{figure}[ht!]
   \centering
      \includegraphics{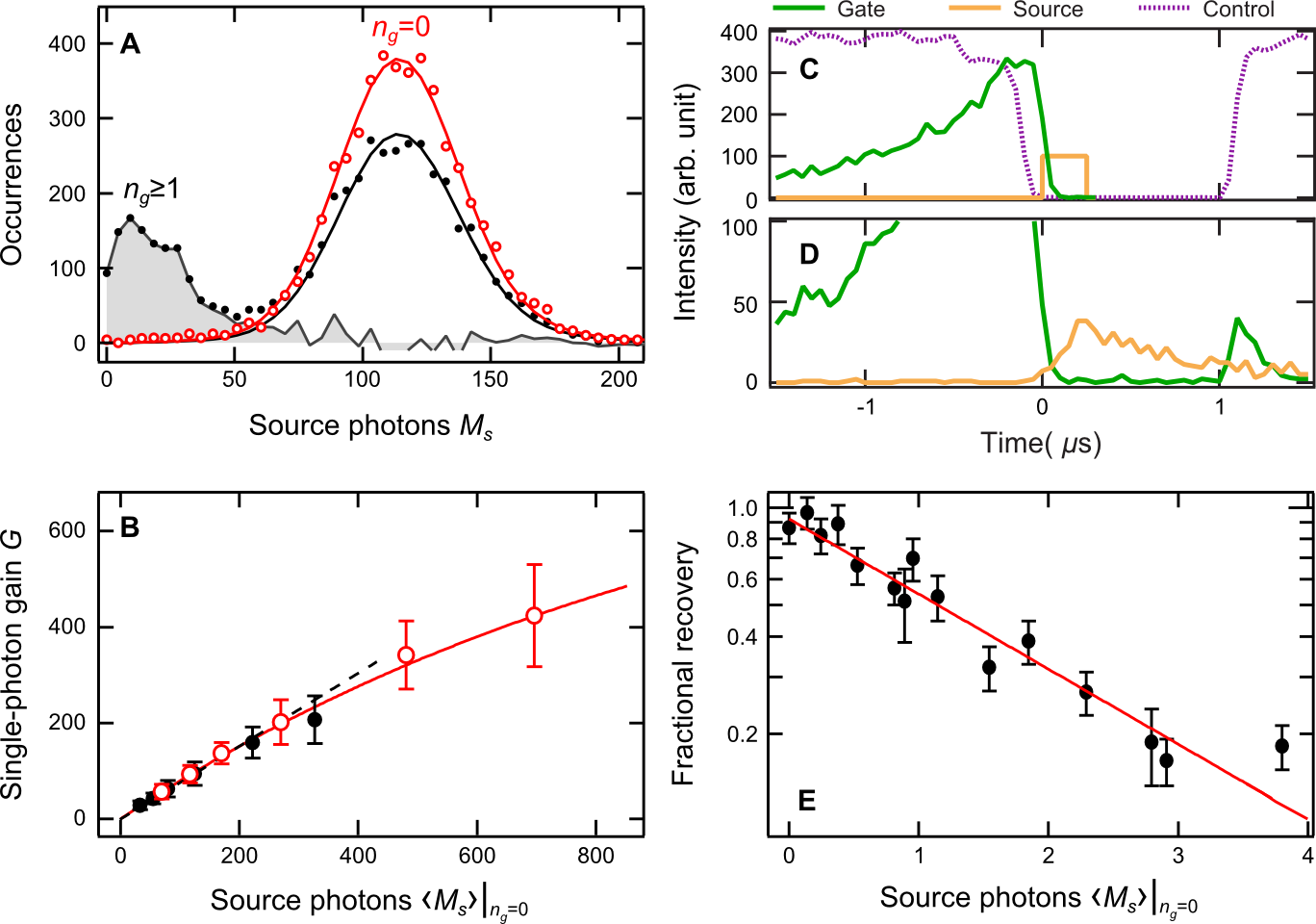}
   \caption{{\bf Fig. 4. Measurement of transistor gain.} (A) Histogram of the integrated source photon number $M_s$ in a $50~\mu$s window. The graph shows $M_s$ for no applied gate photon ($n_g=0$, open red circles) with a Poissonian fit and for a coherent state with $\aver{n_g}=0.4$ stored gate photons (solid black circles). The gray area indicates the contribution from events with $n_g\geq1$, with average value denoted by $\aver{M_s}\big{|}_{n_g\geq1}$ . (B) Transistor gain $G=\aver{M_s}\big{|}_{n_g=0} -\aver{M_s}\big{|}_{n_g\geq1}$ as a function of source strength $\aver{M_s}\big{|}_{n_g=0}$ for integration times of $25~\mu$s (solid black circles) and $50~\mu$s (open red circles), with a linear fit to the first 9 data points (black dashed line) and with exponential fit for gain saturation (red line). {\bf Timing sequence for retrieval operation} with input pulses (C) and output pulses (D). (The actual gate, control, and source beam waveforms are shown, but relative powers are not to scale.) First, the control beam is adiabatically ramped down at $t = 0$ to store a gate photon in the atomic medium. Then a source pulse is sent onto the cavity and its transmission measured. Subsequently, the control beam is adiabatically ramped up to retrieve and detect the gate photon. The combined storage and retrieval efficiency in the absence of source light after a storage time of $1~\mu$s is $(3.0\pm0.1)\%$. (E) {\bf Measurement of transistor gain in retrieval mode.} The average fractional retrieval efficiency of the gate photon after $1~\mu$s is plotted vs. $\aver{M_s}\big{|}_{n_g=0}$ with an exponential fit. The fitted source photon number resulting in $e^{-1}$ reduction is $M_{s0}=1.9 \pm 0.1$ outside of the cavity ($M_{s0}=2.8 \pm 0.2$ before outcoupling losses, in good agreement with the theoretical value $2.8 \pm 0.1$).}
\end{figure}

To operate the device with gate retrieval where the stored photon is recovered in the original optical mode after switching the source light, the source integration time is reduced to $1~\mu$s, less than the measured lifetime $\tau=(2.1\pm0.1)~\mu$s of the collective spin excitation. In this case we can directly measure the cavity transmission probability conditioned on the detection of a gate photon, given by the gate-source cross-correlation function $g_{gs}^{(2)}=\aver{n_g n_s}/(\aver{n_g} \aver{n_s})$ in the limit $\aver{n_g}, \aver{n_s} \ll 1$. On cavity resonance we measure $g_{gs}^{(2)}=0.29^{+0.09}_{-0.08}$ for $0.2$ average retrieved gate photons and $0.1$ average source photons transmitted. These average photon values were chosen to minimize the two-photon probability in each beam while ensuring that the signal-to-background ratio remains sufficiently high. If we subtract independently measured detector backgrounds \cite{SOM}, we find a corrected value of $\bar{g}_{gs}^{(2)}=0.17^{+0.08}_{-0.06}$. This substantial anticorrelation, arising from the effective interaction between two initially uncorrelated photons of different wavelengths, is in good agreement with the value $T=0.09 \pm 0.01$ deduced from Fig.~3B, and the value $T=0.16\pm0.06$ expected from first principles, as discussed below.

Finally, we determine the available gain $G_r$ in retrieval mode by measuring the retrieval reduction as a function of source photon number, and display the result in Fig.~4E. In the process, it is only the scattering of a source photon into free space that reveals the location of the excited atom, thereby collapsing the collective state into a single-atom state, and preventing the retrieval. This scattering is suppressed in the strong-coupling limit of cavity QED, as discussed below, and the observed dependence of retrieval on source photon number agrees well with the theoretical model.  The physical gain of the device operated at $1/e$ retrieval reduction is $G_r=2.2 \pm 0.2$, and the available gain outside the cavity is $1.4 \pm 0.1$ (lower due to the 0.66 outcoupling efficiency). This demonstrates a gain exceeding unity in transistor operation where the gate photon is preserved.

Our observations can be quantitatively understood in a simple cavity QED model: One atom in state $\ket{s}$ reduces the cavity transmission \cite{Birnbaum05,Kubanek08} by a factor $T = (1+\eta)^{-2}$, where $\eta$ is the single-atom cooperativity \cite{Tanji-Suzuki11a}. In the strong-coupling regime of cavity QED, $\eta \gg 1$, already one stored gate photon can thus switch the source beam from transmission to reflection with high contrast. The cooperativity parameter $\eta$ also governs the number of source photons that can be switched: the destruction probability for the collective excitation is given by the probability of scattering a photon into free space on the $\ket{s} \rightarrow \ket{e}$ transition. Such scattering probability is suppressed by cavity to $2\eta/(1+\eta)^2$ in the regime of continuous cavity excitation \cite{Tanji-Suzuki11a}. For $\eta \gg 1$ high transistor gain can be achieved, and the gate photon can be still retrieved from the atomic ensemble afterwards. Note that, as the cavity blocking mechanism does not rely on the collective nature of the atomic excitation, even when the latter is destroyed, the remaining atom in state $\ket{s}$ continues to switch the source beam, leading to high gain $G \gg 1$ in the incoherent regime.

For the present system \cite{Tanji-Suzuki11}, the cooperativity for a two-level atom at an antinode is $\eta_0=8.6 \pm 0.4$. Averaging over polarization factors, the cavity standing wave, and the gate beam reduces the available cooperativity. The directly averaged cooperativity value is $\aver{\eta}=2.8$, while the effective cooperativities for the transmission extinction and the attenuation photon number are $\bar{\eta}_{T}=1.5$, and $\bar{\eta}_{a}=3.3$, respectively \cite{SOM}. The theoretical model is in agreement with our measurements of the transmission reduction induced by one gate photon, and with the measured dependence of gate photon retrieval efficiency on source photon number, as displayed in Fig.~4E. The theoretical model, after including optical pumping into other magnetic sublevels \cite{SOM}, also reproduces the measured cavity transmission histogram, as shown in Fig.~3D.

Our system constitutes a testbed in which we have explored the physical principles relevant to an all-optical transistor based on cavity QED with an atomic ensemble. Before it can be used as a practical device, it will be necessary to improve the input and output coupling efficiencies for the gate and source photons, which limit the usable gain in the system. The combined storage and retrieval efficiency of 3$\%$ for the gate photon is limited primarily by the optical density. The latter could be improved by using a deeper trap, in combination with further cooling of the atomic ensemble, which would also increase the gate photon storage time that is currently limited by Doppler broadening. The cavity outcoupling efficiency for the source photons of 0.66 could be improved to 0.97 by using state-of-the-art mirrors \cite{Birnbaum05,Brennecke07,Kubanek08}.

The present work opens up new perspectives for all-optical information processing with strong deterministic interactions between initially uncorrelated, distinguishable photons. The gain $G_r>1 $ in operation with gate photon retrieval may enable not only hitherto unexplored all-optical quantum circuits with feedback and gain, but also the non-destructive detection of the gate photon, a feat that has so far only been accomplished for microwave photons confined in a cavity \cite{Guerlin07}. We further note that the correlations between one gate and multiple source photons produced by the effective photon-photon interaction can be used to create two-mode entangled states of many photons. Finally, cavities with larger cooperativity \cite{Birnbaum05,Brennecke07,Colombe07,Kubanek08}, may enable high-fidelity deterministic photonic quantum gates.

\bibliography{arXiv_SinglePhotonTransistor}

\begin{thebibliography}{10}

\bibitem{Birnbaum05}
K.~M. Birnbaum, {\it et~al.\/}, {\it Nature\/} {\bf 436}, 87 (2005).

\bibitem{Brennecke07}
F.~Brennecke, {\it et~al.\/}, {\it Nature\/} {\bf 450}, 268 (2007).

\bibitem{Colombe07}
Y.~Colombe, {\it et~al.\/}, {\it Nature\/} {\bf 450}, 272 (2007).

\bibitem{Kubanek08}
A.~Kubanek, {\it et~al.\/}, {\it Phys. Rev. Lett.\/} {\bf 101}, 203602 (2008).

\bibitem{Tanji-Suzuki11}
H.~Tanji-Suzuki, W.~Chen, R.~Landig, J.~Simon, V.~Vuleti\'{c}, {\it Science\/}
  {\bf 333}, 1266 (2011).

\bibitem{Brooks12}
D.~Brooks, {\it et~al.\/}, {\it Nature\/} {\bf 488}, 476 (2012).

\bibitem{Michler00}
P.~Michler, {\it et~al.\/}, {\it Science\/} {\bf 290}, 2282 (2000).

\bibitem{Press07}
D.~Press, {\it et~al.\/}, {\it Phys. Rev. Lett.\/} {\bf 98}, 117402 (2007).

\bibitem{Fushman08}
I.~Fushman, {\it et~al.\/}, {\it Science\/} {\bf 320}, 769 (2008).

\bibitem{Volz12}
T.~Volz, {\it et~al.\/}, {\it Nat. Photon.\/} {\bf 6}, 605 (2012).

\bibitem{Bose12}
R.~Bose, D.~Sridharan, H.~Kim, G.~S. Solomon, E.~Waks, {\it Phys. Rev. Lett.\/}
  {\bf 108}, 227402 (2012).

\bibitem{Dudin12}
Y.~O. Dudin, A.~Kuzmich, {\it Science\/} {\bf 336}, 887 (2012).

\bibitem{Peyronel12}
T.~Peyronel, {\it et~al.\/}, {\it Nature\/} {\bf 488}, 57 (2012).

\bibitem{Hwang09}
J.~Hwang, {\it et~al.\/}, {\it Nature\/} {\bf 460}, 76 (2009).

\bibitem{Bajcsy09}
M.~Bajcsy, {\it et~al.\/}, {\it Phys. Rev. Lett.\/} {\bf 102}, 203902 (2009).

\bibitem{Lo11}
H.-Y. Lo, {\it et~al.\/}, {\it Phys. Rev. A\/} {\bf 83}, 041804 (2011).

\bibitem{Schmidt96}
H.~Schmidt, A.~Imamo\v{g}lu, {\it Opt. Lett.\/} {\bf 21}, 1936 (1996).

\bibitem{Imamoglu97}
A.~Imamo\ifmmode~\bar{g}\else \={g}\fi{}lu, H.~Schmidt, G.~Woods, M.~Deutsch,
  {\it Phys. Rev. Lett.\/} {\bf 79}, 1467 (1997).

\bibitem{Harris98}
S.~Harris, Y.~Yamamoto, {\it Phys. Rev. Lett.\/} {\bf 81}, 3611 (1998).

\bibitem{Fleischhauer00}
M.~Fleischhauer, M.~D. Lukin, {\it Phys. Rev. Lett.\/} {\bf 84}, 5094  (2000).

\bibitem{Fleischhauer05}
M.~Fleischhauer, A.~Imamoglu, J.~Marangos, {\it Rev. Mod. Phys.\/} {\bf 77},
  633 (2005).

\bibitem{Dayan08}
B.~Dayan, {\it et~al.\/}, {\it Science\/} {\bf 319}, 1062 (2008).

\bibitem{Turchette95}
Q.~A. Turchette, C.~J. Hood, W.~Lange, H.~Mabuchi, H.~J. Kimble, {\it Phys.\
  Rev.\ Lett.\/} {\bf 75}, 4710 (1995).

\bibitem{Thompson92}
R.~J. Thompson, G.~Rempe, H.~J. Kimble, {\it Phys.\ Rev.\ Lett.\/} {\bf 68},
  1132 (1992).

\bibitem{Chang07}
D.~E. Chang, A.~S. Sorensen, E.~A. Demler, M.~D. Lukin, {\it Nat. Phys.\/} {\bf
  3}, 807 (2007).

\bibitem{Liu01}
C.~Liu, Z.~Dutton, C.~H. Behroozi, L.~Hau, {\it Nature\/} {\bf 409}, 490
  (2001).

\bibitem{Phillips01}
D.~F. Phillips, A.~Fleischhauer, A.~Mair, R.~L. Walsworth, M.~D. Lukin, {\it
  Phys.\ Rev.\ Lett.\/} {\bf 86}, 783 (2001).

\bibitem{Gorshkov2007}
A.~V. Gorshkov, A.~Andr\'e, M.~Fleischhauer, A.~S. S\o{}rensen, M.~D. Lukin,
  {\it Phys. Rev. Lett.\/} {\bf 98}, 123601 (2007).

\bibitem{SOM}
Materials and methods are available as supporting material on {\it Science}
  Online.

\bibitem{Tanji-Suzuki11a}
H.~Tanji-Suzuki, {\it et~al.\/}, {\it Adv. At. Mol Opt.\/} {\bf 60}, 201
  (2011).

\bibitem{Guerlin07}
C.~Guerlin, {\it et~al.\/}, {\it Nature\/} {\bf 448}, 889 (2007).

\end{thebibliography}

\bibliographystyle{science}

% Following is a new environment, {scilastnote}, that's defined in the
% preamble and that allows authors to add a reference at the end of the
% list that's not signaled in the text; such references are used in
% *Science* for acknowledgments of funding, help, etc.

\begin{scilastnote}
\item[] \textbf{Acknowledgements:} This work was supported by the NSF and AFOSR.  K.B. gratefully acknowledges support from the National Science Foundation (NSF) through the Graduate Research Fellowship (0645960).  R.B. gratefully acknowledges support from FWF doctoral program CoQuS (W1210).
\end{scilastnote}

\clearpage


\section*{Supplementary material}

# Supplementary Materials for

## All-Optical Switch and Transistor Gated by One Stored Photon

Wenlan Chen, Kristin M. Beck, Robert Bücker, Michael Gullans, Mikhail D. Lukin,
Haruka Tanji-Suzuki, Vladan Vuletić

correspondence to:  vuletic@mit.edu

**This PDF file includes:**

    Materials and Methods
    Supplementary Text
    Fig. S1
    Table S1
    Full Reference List

**Materials and Methods**

To prepare the atomic ensemble, a magneto-optical trap (MOT) of $^{133}$Cs atoms is overlapped with a far-off-resonant optical-lattice trap of trap depth $U_0/h$=6.9 MHz operated at 937 nm in the TEM$_{00}$ mode of the 1.4 cm-long cavity. We compress the MOT for 27 ms using a magnetic field gradient of 22 G/cm while cooling the atoms in an optical molasses, thus loading typically $5 \times 10^5$ Cs atoms into the optical-lattice trap at a temperature of 35 μK. To reduce cavity blocking and shifts due to atoms not addressed by the gate beam, we remove atoms outside this beam by spatially-resolved shelving of atoms into the hyperfine manifold $F$=3 and then pushing excess atoms out of the trap by radiation pressure on the $|6S_{1/2}, F=4\rangle \rightarrow |6P_{3/2}, F'=5\rangle$ transition. The remaining $1.6 \times 10^4$ atoms are then optically pumped into state $|g\rangle$ along the quantization axis ($x$ axis), defined by a 12 G magnetic field (Fig. 1A), resulting in a typical resonant optical depth $OD$=0.9 for the $\sigma^+$-polarized gate beam propagating along the $x$ axis. The π-polarized control beam is incident along the $y$ axis with a variable Rabi frequency up to $\Omega=2\pi \times 1.5$ MHz.

**Supplementary Text**

<u>Detection calibration</u>

The gain measurements depend on a careful calibration of the detection efficiency for the source (cavity) mode. We use a PerkinElmer single photon counting module SPCM-AQR-14-FC that has a rate-dependent efficiency $E(R)$ that is well approximated as

$$E(R)=0.45\times(1-3.4\times10^{-3}-6.5\times10^{-5}R) \quad (1)$$

where 0.45 is the module photon detection efficiency and $R$ is the photon rate in kCount/s. The functional form is determined by a fit to detector linearity measurements provided by the company for the detector. The real cavity output rate is then

$$R_{out} = \frac{R-b}{E(R)\cdot T} \quad (2)$$

where $b$ is the independently measured background count rate and $T$ is the measured optical transmission of the detection path so that the path detection efficiency is $DE_s=E(R)\cdot T$. The detector efficiency modification only makes an appreciable difference in Figs. 4B, S1C and S1D. For the highest rates, approximately 130 counts in 50 μs or 2600 kCounts/s, the detector efficiency $E(R)\approx 0.38$, a ~20% decrease in the detection efficiency.

The background count rate $b$ is determined for each measurement and is less than 1 kCount/s for the data presented here. These counts come from trap laser photons, stray room light and detector dark counts.

The transmitted source photon number $M_s$ is equal to the photon output rate by the cavity, simply the measured rate divided by the path detection efficiency for continuous excitation. (For pulsed excitation, as in the measurement of the attenuation photon number in Fig. 4E, the output photon rate is multiplied by 2 because an undriven symmetric cavity decays equally out of each side). Equivalently, $M_s = \frac{T}{T+L}\int dt\, m_c(t)\kappa$ as defined in the text, where $m_c(t)$ is the intracavity photon number at time $t$, $\kappa$ is the

2

cavity linewidth, and the ratio $\frac{\mathcal{T}}{\mathcal{T}+\mathcal{L}}$ =0.66 is the cavity outcoupling efficiency with mirror transmission $\mathcal{T}$=27ppm and mirror loss $\mathcal{L}$=14ppm.

Spatial averaging

There are three separate spatially varying factors that enter our theoretical predictions for the one-atom cooperativity, and quantities derived from it. First, the weak source mode is a standing wave in the cavity. This standing wave modulates the cavity-atom interaction strength, resulting in a position-dependent cooperativity $\eta \cos^2(kz)$ with wavevector $k=2\pi/852$ nm. The off-resonant dipole trap also forms a standing wave in the cavity. Atoms are confined near to the antinodes of this trap, resulting in a (937 nm)/2≈469 nm-periodic atomic density function $n(z)$, which we approximate as an array of Gaussian wavepackets with a width of 54 nm, calculated from the temperature, the trap depth $U_0/h$ and the trap wavelength. Finally, the location of the excited atom(s) in the ensemble is also spatially dependent because the gate mode is tightly focused to a waist of $w=(2.2 \pm 0.5)$ μm in the atomic ensemble.

The predicted values of the cavity transmission $T$ and atomic scattering probability into free space $S$ for a single atom are calculated from the weighted average

$$\{f\} = \int dz\, n(z) e^{-2z^2/w^2} f(\eta \cos^2(kz)) \Big/ \int dz\, n(z) e^{-2z^2/w^2} \qquad (3)$$

$f(x)$ is replaced by $1/(1+x)^2$ for the transmission and by $2x/(1+x)^2$ for the scattering *(30)*. The predicted cross-correlation function is $g^{(2)}_{gs} = T^{-1}$, and the predicted attenuation photon number is $\mathrm{M}_{s0} = S^{-1}$.

The effective cooperativity for transmission extinction $\bar{\eta}_T$ is the cooperativity that satisfies the equation

$$\{(1+x)^{-2}\} = (1+\bar{\eta}_T)^{-2} \qquad (4)$$

Similarly, the effective cooperativity for attenuation photon number is

$$\{2x(1+x)^{-2}\} = 2\bar{\eta}_a(1+\bar{\eta}_a)^{-2} \qquad (5)$$

Theoretical prediction for average cavity transmission

The average transmission of an optical cavity that is coupled to $N$ stored photons with cooperativity $\eta$ is

$$T(\delta, N) = \left[\left(1 + \frac{N\eta}{1+(2\delta/\kappa)^2}\right)^2 + (2\delta/\kappa)^2\left(1 - \frac{\kappa}{\Gamma}\frac{N\eta}{1+(2\delta/\kappa)^2}\right)^2\right]^{-1} \qquad (6)$$

where $\delta$ is the source-cavity detuning, $\kappa$ is the cavity linewidth and $\Gamma$ is the atomic linewidth. If we store a coherent gate pulse with mean photon number $\langle n_g \rangle$ in the ensemble, the predicted transmission is

$$\sum_N p_N \{T(\delta, N)\} = e^{-\langle n_g \rangle} \sum_N \frac{\langle n_g \rangle^N}{N!} \{T(\delta, N)\} \qquad (7)$$

3

where $p_N$ describes the distribution of number states $|N\rangle$ in the coherent state and the average $\{T(\delta, N)\}$ accounts for the spatially varying cooperativity as described in the previous section.

Theoretical prediction for histogrammed cavity transmission

The probability of detecting a fixed transmitted source photon number $s$ at a given source-cavity detuning $\delta$ for an optical cavity that is coupled to $N$ stored photons with cooperativity $\eta$ is

$$P(\delta, N, s) = e^{-T(\delta, N)} \frac{T(\delta, N)^s}{s!} \qquad (8)$$

That is, the transmission is Poisson distributed (due to shot noise) with an average value given by the transmission in Eq. (6).

If we store a coherent gate pulse in the ensemble, the predicted probability is

$$\sum_N p_N \{P(\delta, N, s)\} = e^{-\langle n_g \rangle} \sum_N \frac{\langle n_g \rangle^N}{N!} \{P(\delta, N, s)\} \qquad (9)$$

where $p_N$ describes the distribution of number states $|N\rangle$ in the coherent gate pulse and the average $\{P(\delta, N, s)\}$ accounts for the spatially varying cooperativity.

Optical pumping and saturation

At high photon number, the transistor gain saturates (Fig. 4B). This comes from two effects: optical pumping of atoms originally in state $|s\rangle$ into magnetic sublevels with weaker coupling to the cavity, and atomic saturation. This effect is clearly visible in the resonant transmission histograms (Fig. S1), where the peak extinction factor $T^{-1}$ decreases from approximately 17 to 2 as the source power increases (see Fig. S1). To account for this effect, the prediction in Fig. 2 uses the reduced coupling extracted from the corresponding resonant transmission histogram for those data.

Background correction of the cross correlation function

Measured $g^{(2)}_{gs}$ functions differ from their ideal values as a result of detector backgrounds. Accounting for this effect by expressing the measured (i.e. retrieved or transmitted) photon rates as the sum of the signal $n_i$ and the uncorrelated background $b_i$ in each mode, $i=g,s$, we have:

$$g^{(2)}_{gs} = \frac{\langle (n_g + b_g)(n_s + b_s) \rangle}{\langle n_g + b_g \rangle \langle n_s + b_s \rangle} = \frac{\langle n_g n_s \rangle + \langle b_g \rangle \langle n_s \rangle + \langle n_g \rangle \langle b_s \rangle + \langle b_g \rangle \langle b_s \rangle}{\langle n_g \rangle \langle n_s \rangle + \langle b_g \rangle \langle n_s \rangle + \langle n_g \rangle \langle b_s \rangle + \langle b_g \rangle \langle b_s \rangle} \qquad (10)$$

Dividing through by $\langle n_g \rangle \langle n_s \rangle$, identifying $\overline{g}^{(2)}_{gs} = \langle n_g n_s \rangle / (\langle n_g \rangle \langle n_s \rangle)$ and gathering background-to-signal ratios into a single term $B = \langle b_s \rangle / \langle n_s \rangle + \langle b_g \rangle / \langle n_g \rangle + \langle b_g \rangle \langle b_s \rangle / (\langle n_g \rangle \langle n_s \rangle)$, we obtain

$$g^{(2)}_{gs} = \frac{\overline{g}^{(2)}_{gs} + B}{1 + B} \qquad (11)$$

4

which can be inverted to find the background-corrected $\overline{g}_{gs}^{(2)}$ from measured values for $B$ and $g_{gs}^{(2)}$

$$\overline{g}_{gs}^{(2)} = (1+B)g_{gs}^{(2)} - B \tag{12}$$

**Fig. S1.**

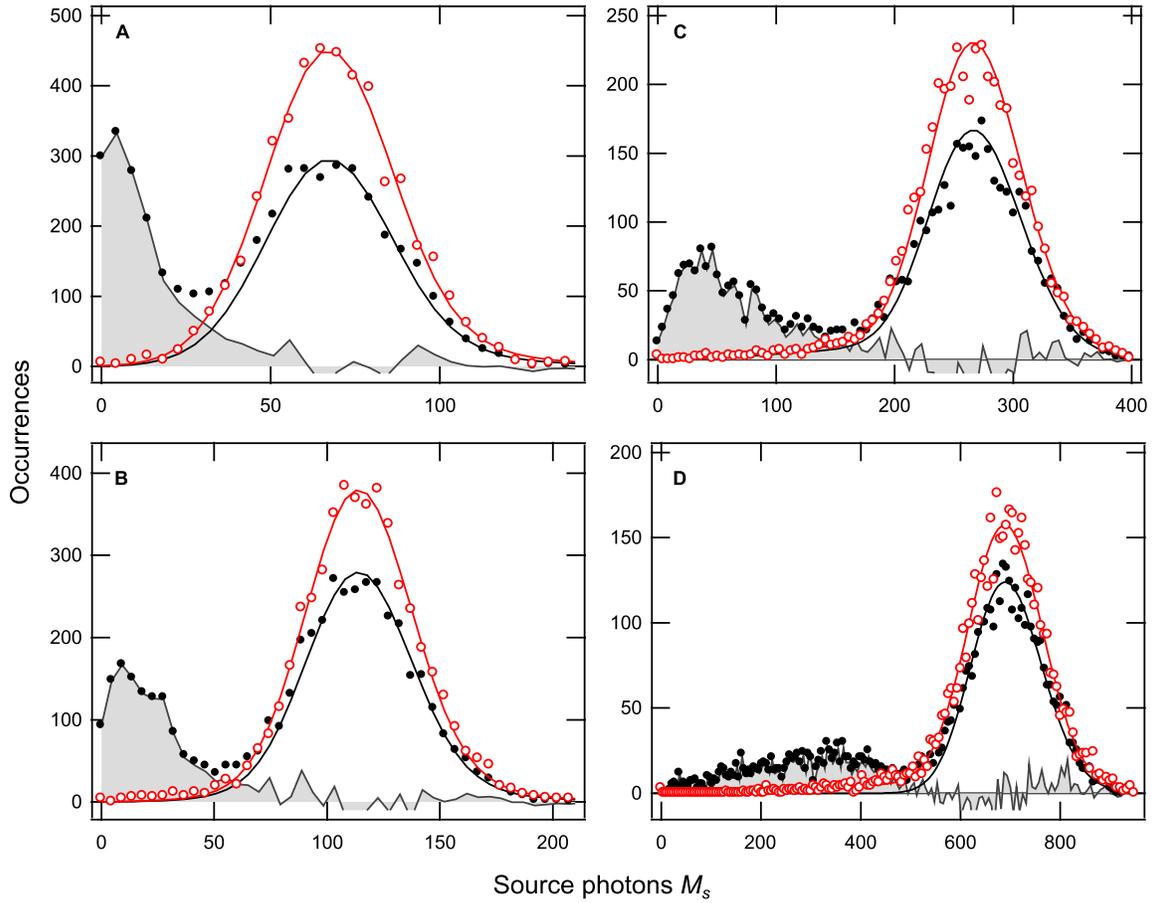

Histograms of $M_s$ in a 50 μs integration window. The average detected photon counts for the empty cavity were (A) 15, (B) 25, (C) 55, and (D) 130, corresponding to $\langle M_s \rangle|_{n_g=0}$ = (A) 70, (B) 120, (C) 270 and (D) 700. Each graph shows cavity transmission for $n_g=0$ (open red circles) and cavity transmission with $\langle n_g \rangle \approx 0.4$ (solid black circles). The grayed area indicates the contribution to the cavity transmission from $\langle n_g \rangle \geq 1$ determined by subtracting a Poissonian curve and a pedestal that are fit to the $\langle n_g \rangle = 0$ curve from the $\langle n_g \rangle \approx 0.4$ cavity transmission. The average value of this area up to $\sim \langle M_s \rangle|_{n_g=0}$ is denoted $\langle M_s \rangle|_{n_g \geq 1}$. Note that (B) presents the same histograms as Fig. 4A.

**Table S1.**

| Cavity | | $\lambda=852$nm |
|---|---|---|
| Mirror separation | $L$ | 13.7 mm* |
| Mirror curvature radius | $R_M$ | $(10.0 \pm 0.2)$ mm |
| Mirror transmission | $\mathcal{T}$ | 27 ppm |
| Mirror loss | $\mathcal{L}$ | 14 ppm |
| Free spectral range | $\omega_{FSR}/(2\pi)$ | 10909 MHz* |
| Linewidth | $\kappa/(2\pi)$ | $(142 \pm 1)$ kHz |
| Finesse | $\mathcal{F}_{852}$ | $(7.71 \pm 0.05) \times 10^4$ |
| Mode waist | $w_{852}$ | $(35.5 \pm 0.2)$ μm |
| Optical dipole trap | | $\lambda=937$nm |
| Finesse | $\mathcal{F}_{937}$ | $(37.2 \pm 0.2)$ μm |
| Mode waist | $w_{937}$ | $368 \pm 20$ |
| Trap depth | $U_0/h$ | $(6.9 \pm 0.4)$ MHz |
| Gate (free-space) path | | |
| Beam waist | $w_g$ | $(2.2 \pm 0.5)$ μm |
| Oscillator strength | $f_{gd}$ | 0.42 |
| Optical density | $OD$ | $0.9 \pm 0.1$ |
| Detection efficiency | $DE_g$ | $0.13 \pm 0.01$ |
| Background rate | | $\leq 1$ kHz |
| Source (cavity) path | | |
| Oscillator strength | $f_{se}$ | 0.50 |
| Antinode cooperativity | $\eta$ | $4.3 \pm 0.2$ |
| Detection efficiency | $DE_s$ | $0.22 \pm 0.02$ |
| Background rate | | $\leq 1$ kHz |

Experimental parameters. The mirror separation and cavity free spectral range (marked *) have standard deviations better than the precision we quote.

**Full Reference List**

1. K. M. Birnbaum, *et al.*, *Nature*, **436**, 87 (2005).
2. F. Brennecke, *et al.*, *Nature* **450**, 268 (2007).
3. Y. Colombe, *et al.*, *Nature* **450**, 272 (2007).
4. A. Kubanek, *et al.*, *Phys. Rev. Lett.* **101**, 203602 (2008).
5. H. Tanji-Suzuki, W. Chen, R. Landig, J. Simon, V. Vuletić, *Science* **333**, 1266 (2011).
6. D. Brooks, *et al.*, *Nature* **488**, 476 (2012).
7. P. Michler, *et al.*, *Science* **290**, 2282 (2000).
8. D. Press, *et al.*, *Phys. Rev. Lett.* **98**, 117402 (2007).
9. I. Fushman, *et al.*, *Science* **320**, 769 (2008).
10. T. Volz, *et al.*, *Nat. Photon.* **6**, 605 (2012).
11. R. Bose, D. Sridharan, H. Kim, G. S. Solomon, E. Waks, *Phys. Rev. Lett.* **108**, 227402 (2012).
12. Y. O. Dudin, A. Kuzmich, *Science* **336**, 887 (2012).
13. T. Peyronel, *et al.*, *Nature* **488**, 57 (2012).
14. J. Hwang, *et al.*, *Nature* **460**, 76 (2009).
15. M. Bajcsy, *et al.*, *Phys. Rev. Lett.* **102**, 203902 (2009).
16. H.-Y. Lo, *et al.*, *Phys. Rev. A* **83**, 041804 (2011).
17. H. Schmidt, A. Imamoğlu, *Opt. Lett.* 21, 1936 (1996).
18. A. Imamoḡlu, H. Schmidt, G. Woods, M. Deutsch, *Phys. Rev. Lett.* **79**, 1467 (1997).
19. S. Harris, Y. Yamamoto, *Phys. Rev. Lett.* **81**, 3611 (1998).
20. M. Fleischhauer, M. D. Lukin, *Phys. Rev. Lett.* **84**, 5094 (2000).
21. M. Fleischhauer, A. Imamoglu, J. Marangos, *Rev. Mod. Phys.* **77**, 633 (2005).
22. B. Dayan, *et al.*, *Science* **319**, 1062 (2008).
23. Q. A. Turchette, C. J. Hood, W. Lange, H. Mabuchi, H. J. Kimble, *Phys. Rev. Lett.* 75, 4710 (1995).
24. R. J. Thompson, G. Rempe, H. J. Kimble, *Phys. Rev. Lett.* **68**, 1132 (1992).
25. D. E. Chang, A. S. Sorensen, E. A. Demler, M. D. Lukin, *Nat. Phys.* **3**, 807 (2007).
26. C. Liu, Z. Dutton, C. H. Behroozi, L. Hau, *Nature* **409**, 490 (2001).
27. A. V. Gorshkov, A. André, M. Fleischhauer, A. S. Sørensen, M. D. Lukin, *Phys. Rev. Lett.* **98**, 123601 (2007).
28. D. F. Phillips, A. Fleischhauer, A. Mair, R. L. Walsworth, M. D. Lukin, *Phys. Rev. Lett.* **86**, 783 (2001).
29. Materials and methods are available as supporting material on *Science* online.
30. H. Tanji-Suzuki, *et al.*, *Adv. At. Mol Opt.* **60**, 201 (2011).
31. C. Guerlin, *et al.*, *Nature* **448**, 889 (2007).



\end{document}